\documentstyle[12pt]{article}

\catcode`\@=11
\def\@citex[#1]#2{%
\if@filesw \immediate \write \@auxout {\string \citation {#2}}\fi
\@tempcntb\m@ne \let\@h@ld\relax \def\@citea{}%
\@cite{%
  \@for \@citeb:=#2\do {%
    \@ifundefined {b@\@citeb}%
      {\@h@ld\@citea\@tempcntb\m@ne{\bf ?}%
      \@warning {Citation `\@citeb ' on page \thepage \space undefined}}%
      {\@tempcnta\@tempcntb \advance\@tempcnta\@ne%
      \@tempcntb\number\csname b@\@citeb \endcsname \relax%
      \ifnum\@tempcnta=\@tempcntb 
        \ifx\@h@ld\relax%
          \edef \@h@ld{\@citea\csname b@\@citeb\endcsname}%
        \else%
          \edef\@h@ld{\ifmmode{-}\else--\fi\csname b@\@citeb\endcsname}%
        \fi%
      \else
        \@h@ld\@citea\csname b@\@citeb \endcsname%
        \let\@h@ld\relax%
      \fi}%
    \def\@citea{,\penalty\@highpenalty\,}%
  }\@h@ld
}{#1}}

\def\@citeb#1#2{{[#1]\if@tempswa , #2\fi}}
\def\@citeu#1#2{{$^{#1}$\if@tempswa , #2\fi }}
\def\@citep#1#2{{#1\if@tempswa , #2\fi}}

\def\bcites{         
        \catcode`\@=11
        \let\@cite=\@citeb
        \catcode`\@=12
}

\def\upcites{         
        \catcode`\@=11
        \let\@cite=\@citeu
        \catcode`\@=12
}

\def\plaincites{      
        \catcode`\@=11
        \let\@cite=\@citep
        \catcode`\@=12
}

\newcount\hour
\newcount\minute
\newtoks\amorpm
\hour=\time\divide\hour by 60
\minute=\time{\multiply\hour by 60 \global\advance\minute by-\hour}
\edef\standardtime{{\ifnum\hour<12 \global\amorpm={am}%
        \else\global\amorpm={pm}\advance\hour by-12 \fi
        \ifnum\hour=0 \hour=12 \fi
        \number\hour:\ifnum\minute<10 0\fi\number\minute\the\amorpm}}
\edef\militarytime{\number\hour:\ifnum\minute<10 0\fi\number\minute}

\def\draftlabel#1{{\@bsphack\if@filesw {\let\thepage\relax
   \xdef\@gtempa{\write\@auxout{\string
      \newlabel{#1}{{\@currentlabel}{\thepage}}}}}\@gtempa
   \if@nobreak \ifvmode\nobreak\fi\fi\fi\@esphack}
        \gdef\@eqnlabel{#1}}
\def\@eqnlabel{}
\def\@vacuum{}
\def\marginnote#1{}
\def\draftmarginnote#1{\marginpar{\raggedright\scriptsize\tt#1}}
\def\draft{
        \pagestyle{plain}
        \overfullrule=2pt
        \oddsidemargin -.5truein
        \def\@oddhead{\sl \phantom{\today\quad\militarytime} \hfil
        \smash{\Large\sl DRAFT} \hfil \today\quad\militarytime}
        \let\@evenhead\@oddhead
        \let\label=\draftlabel
        \let\marginnote=\draftmarginnote
        \def\ps@empty{\let\@mkboth\@gobbletwo
        \def\@oddfoot{\hfil \smash{\Large\sl DRAFT} \hfil}
        \let\@evenfoot\@oddhead}
        \def\@eqnnum{(\theequation)\rlap{\kern\marginparsep\tt\@eqnlabel}%
        \global\let\@eqnlabel\@vacuum}  }

\def\blackfonts{
        \font\blackboard=msbm10 scaled\magstep1
        \font\blackboards=msbm8
        \font\blackboardss=msbm6
}

\def\prep{         
        \catcode`\@=11
        \input art10.sty
        \catcode`\@=12
        
        \let\small\null
        \def\blackfonts{
                \font\blackboard=msbm10
                \font\blackboards=msbm7
                \font\blackboardss=msbm5
        }
        \let\sl\it
        \twocolumn
        \sloppy
        \voffset=-2.54truecm
        \hoffset=-2.54truecm
        \flushbottom
        \parindent 1em
        \leftmargini 2em
        \leftmarginv .5em
        \leftmarginvi .5em
        \marginparwidth 48pt
        \marginparsep 10pt
        \setlength{\columnsep}{2truecm}
        \setlength{\textwidth}{25.4truecm}
        \setlength{\textheight}{17truecm}
        \baselineskip=16pt
        \oddsidemargin .18truein
        \evensidemargin .17truein
}

\def\eqalign#1{\null\,\vcenter{\openup\jot\m@th
  \ialign{\strut\hfil$\displaystyle{##}$&$\displaystyle{{}##}$\hfil
      \crcr#1\crcr}}\,}
\def\eqalignno#1{\displ@y \tabskip\centering
  \halign to\displaywidth{\hfil$\@lign\displaystyle{##}$\tabskip\z@skip
    &$\@lign\displaystyle{{}##}$\hfil\tabskip\centering
    &\llap{$\@lign##$}\tabskip\z@skip\crcr
    #1\crcr}}

\def\section{\@startsection {section}{1}{\z@}{3.ex plus 1ex minus
 .2ex}{2.ex plus .2ex}{\large\bf}}
\def\subsection{\@startsection{subsection}{2}{\z@}{2.75ex plus 1ex minus
 .2ex}{1.5ex plus .2ex}{\bf}}        

\def\appendix{{\newpage\section*{Appendix}}\let\appendix\section%
        {\setcounter{section}{0}
        \gdef\thesection{\Alph{section}}}\section}

\def\abstract{\if@twocolumn
\section*{Abstract}
\else 
\begin{center}
{\bf Abstract\vspace{-.5em}\vspace{0pt}}
\end{center}
\quotation
\fi}

\catcode`\@=12

\def\sqr#1#2{{\vcenter{\vbox{\hrule height.#2pt\hbox{\vrule width.#2pt 
height#1pt \kern#1pt \vrule width.#2pt}\hrule height.#2pt}}}}

\def\=d{\,{\buildrel\rm def\over =}\,}

\def\i3p{\p32\int d^3p}

\def\As{A\hbox to 1pt{\hss /}}
\def\np4{\int d^4p_1\cdots d^4p_{n-1}\, }

\def\Tr{{\rm Tr}\, }

\def\nx4{\int d^4x_1\ldots d^4x_n\, }

\def\kon#1#2{\vbox{\halign{##&&##\cr
\lower4pt\hbox{$\scriptscriptstyle\vert$}\hrulefill &
\hrulefill\lower4pt\hbox{$\scriptscriptstyle\vert$}\cr $#1$&
$#2$\cr}}}

\def\konv#1#2#3{\hbox{\vrule height12pt depth-1pt}
\vbox{\hrule height12pt width#1cm depth-11.6pt}
\hbox{\vrule height6.5pt depth-0.5pt}
\vbox{\hrule height11pt width#2cm depth-10.6pt\kern5pt
      \hrule height6.5pt width#2cm depth-6.1pt}
\hbox{\vrule height12pt depth-1pt}
\vbox{\hrule height6.5pt width#3cm depth-6.1pt}
\hbox{\vrule height6.5pt depth-0.5pt}}
\def\konu#1#2#3{\hbox{\vrule height12pt depth-1pt}
\vbox{\hrule height1pt width#1cm depth-0.6pt}
\hbox{\vrule height12pt depth-6.5pt}
\vbox{\hrule height6pt width#2cm depth-5.6pt\kern5pt
      \hrule height1pt width#2cm depth-0.6pt}
\hbox{\vrule height12pt depth-6.5pt}
\vbox{\hrule height1pt width#3cm depth-0.6pt}
\hbox{\vrule height12pt depth-1pt}}

\def\konw#1#2#3{\hbox{\vrule height12pt depth-1pt}
\vbox{\hrule height12pt width#1cm depth-11.6pt}
\hbox{\vrule height6.5pt depth-0.5pt}
\vbox{\hrule height12pt width#2cm depth-11.6pt \kern5pt
      \hrule height6.5pt width#2cm depth-6.1pt}
\hbox{\vrule height6.5pt depth-0.5pt}
\vbox{\hrule height12pt width#3cm depth-11.6pt}
\hbox{\vrule height12pt depth-1pt}}

\def\i{{\rm int}}

\def\e{{\rm ext}}

\def\r{{\rm ret}}

\def\m3{{\mu_1\mu_2\mu_3}}

\def\co{{\rm Com}}

\def\p{{(+)}}

\def\be{\begin{equation}}       \def\eq{\begin{equation}}
\def\ee{\end{equation}}         \def\eqe{\end{equation}}

\def\bea{\begin{eqnarray}}      \def\eqa{\begin{eqnarray}}
\def\ena{\end{eqnarray}}        \def\eea{\end{eqnarray}}
                                \def\eqae{\end{eqnarray}}

\def\ba{\begin{array}}
\def\ea{\end{array}}
\def\unit{1 \hskip-.3em \raise2pt\hbox{$ \scriptstyle |$ } }

\def\e{\epsilon}           
\def\f{\phi}               

\def\i{\iota}

\def\k{\kappa}                    
\def\l{\lambda}
\def\m{\mu}
\def\n{\nu}
  
\def\p{\pi}                
\def\r{\rho}                                     
\def\s{\sigma}                                   
\def\t{\tau}

\def\D{\Delta}

\def\G{\Gamma}

\def\L{\Lambda}

\def\ca{{\cal A}}

\def\co{{\cal O}}

\def\half{{1 \over 2}}

\def\bop#1{\setbox0=\hbox{$#1M$}\mkern1.5mu
        \vbox{\hrule height0pt depth.04\ht0
        \hbox{\vrule width.04\ht0 height.9\ht0 \kern.9\ht0
        \vrule width.04\ht0}\hrule height.04\ht0}\mkern1.5mu}
\def\Box{{\mathpalette\bop{}}}                        
\def\pa{\partial}                              

\def\>{\rangle} 

\def\<{\langle} 
\def\Dsl{D \hskip-.6em \raise1pt\hbox{$ / $ } }

\def\sl#1{\rlap{\hbox{$\mskip 1 mu /$}}#1}
\def\leftrightarrowfill{$\mathsurround=0pt \mathord\leftarrow \mkern-6mu
       \cleaders\hbox{$\mkern-2mu \mathord- \mkern-2mu$}\hfill
       \mkern-6mu \mathord\rightarrow$}
\def\dvec#1{\vbox{\ialign{##\crcr
       \leftrightarrowfill\crcr\noalign{\kern-1pt\nointerlineskip}
       $\hfil\displaystyle{#1}\hfil$\crcr}}}          
\def\hook#1{{\vrule height#1pt width0.4pt depth0pt}}
\def\leftrighthookfill#1{$\mathsurround=0pt \mathord\hook#1
       \hrulefill\mathord\hook#1$}
\def\underhook#1{\vtop{\ialign{##\crcr                 
       $\hfil\displaystyle{#1}\hfil$\crcr
       \noalign{\kern-1pt\nointerlineskip\vskip2pt}
       \leftrighthookfill5\crcr}}}
\def\smallunderhook#1{\vtop{\ialign{##\crcr      
       $\hfil\scriptstyle{#1}\hfil$\crcr
       \noalign{\kern-1pt\nointerlineskip\vskip2pt}
       \leftrighthookfill3\crcr}}}

\def\sfrac#1#2{{\vphantom1\smash{\lower.5ex\hbox{\small$#1$}}\over
       \vphantom1\smash{\raise.4ex\hbox{\small$#2$}}}} 
\def\bfrac#1#2{{\vphantom1\smash{\lower.5ex\hbox{$#1$}}\over
       \vphantom1\smash{\raise.3ex\hbox{$#2$}}}}      
\def\afrac#1#2{{\vphantom1\smash{\lower.5ex\hbox{$#1$}}\over#2}}  
\def\on#1#2{{\buildrel{\mkern2.5mu#1\mkern-2.5mu}\over{#2}}}
\def\ddt#1{\on{\hbox{\LARGE .\kern-2pt.}}#1}             
\def\tdt#1{\on{\hbox{\LARGE .\kern-2pt.\kern-2pt.}}#1}   

\def\boxes#1{
       \newcount\num
       \num=1
       \newdimen\downsy
       \downsy=-1.5ex
       \mskip-2.8mu
       \bo
       \loop
       \ifnum\num<#1
       \llap{\raise\num\downsy\hbox{$\bo$}}
       \advance\num by1
       \repeat}
\def\boxup#1#2{\newcount\numup
       \numup=#1
       \advance\numup by-1
       \newdimen\upsy
       \upsy=.75ex
       \mskip2.8mu
       \raise\numup\upsy\hbox{$#2$}}

\newskip\humongous \humongous=0pt plus 1000pt minus 1000pt
\def\caja{\mathsurround=0pt}
\def\eqalign#1{\,\vcenter{\openup2\jot \caja
       \ialign{\strut \hfil$\displaystyle{##}$&$
       \displaystyle{{}##}$\hfil\crcr#1\crcr}}\,}
\newif\ifdtup

\def\1ov4{{1\over 4}}

\def\Tr{{\rm Tr}}

\def\pa{\partial}

\def\ddt{\dot{\t}}

\def\pa{\partial}

\def\half{{1 \over 2}}

\def\ba{\begin{eqnarray}}
\def\ea{\end{eqnarray}}

\begin{document}
\null\vskip-24pt
\hfill SPIN-1999/23
\vskip-10pt
\hfill {\tt hep-th/9910023}
\vskip0.3truecm
\begin{center}
\vskip 2truecm
{\Large\bf
Quantum effective action from the AdS/CFT correspondence
}\\ 
\vskip 1.5truecm
{\large\bf
Kostas Skenderis\footnote{
email:{\tt K.Skenderis@phys.uu.nl}}
and Sergey N.~Solodukhin\footnote{
email:{\tt S.Solodukhin@phys.uu.nl}}
}\\
\vskip 1truecm
{\it Spinoza Institute, University of Utrecht,\\
Leuvenlaan 4, 3584 CE Utrecht, The Netherlands}
\vskip 1truemm
\end{center}
\vskip 1truecm
\noindent{\bf Abstract:}
We obtain an Einstein metric of constant negative curvature 
given an arbitrary boundary metric in three dimensions, and a 
conformally flat one given an arbitrary conformally flat boundary 
metric in other dimensions. In order to compute the 
on-shell value of the gravitational action for 
these solutions, we propose to integrate the radial coordinate from 
the boundary till a critical value where the bulk 
volume element vanishes. The result, which is a 
functional of the boundary metric, 
provides a sector of the quantum effective action common to all
conformal field theories that have a gravitational 
description. We verify that the so-defined boundary effective action 
is conformally invariant in odd (boundary) dimensions
and has the correct conformal anomaly in even (boundary) dimensions.
In three dimensions and for arbitrary static boundary metric
the bulk metric takes a rather simple form.
We explicitly carry out the computation 
of the corresponding effective action  and find that it
equals the non-local Polyakov action.

\begin{center}
{\it PACS number(s): 11.25.-w,04.62.+v,04.60.-m,11.25.Hf}
\end{center}
\vskip 1cm
\newpage
\baselineskip=.8cm

There is accumulating evidence that certain conformal field theories (CFT)
have a dual description in terms of string (or M) theory on anti-de 
Sitter (adS) spaces \cite{Malda,Gubs,Wit}. This is a strong/weak coupling
duality, so perturbative calculations on one side of the 
correspondence yield strong coupling results on the other side. 
In particular, the strong coupling limit of these CFTs has a description in 
terms of supergravity theory. 

Certain aspects of the correspondence are rather universal 
and apply to all cases that such a duality exists. 
For example, the leading in the large $N$ 
contribution to the Weyl anomaly is universal \cite{HS}. 
More generally, (certain) correlation functions of the energy momentum tensor 
at leading order can also be computed by using only the 
purely gravitational part of the corresponding supergravity
and are therefore universal. 

To obtain this universal behavior one would need to solve 
Einstein's equations with arbitrary Dirichlet boundary conditions
(more precisely, one needs to determine an Einstein manifold
of constant negative curvature given a conformal structure 
at infinity). This is a rather difficult problem whose solution
in all generality has yet to be worked out. 
An existence theorem for such Einstein's metrics 
has been proved by Graham and Lee \cite{GrahamLee} 
for manifolds $X_{d+1}$ that are topologically a $(d{+}1)$-ball, 
so the boundary at infinity is a $d$-sphere $S^d$, and conformal structures 
sufficiently close to the standard one. Moreover, one can 
explicitly obtain an asymptotic expansion of the bulk metric near infinity 
starting from any boundary metric \cite{FG,HS}. 
This information is sufficient 
in order to obtain the counterterms that render the effective
action finite and the conformal anomaly \cite{HS}.
However, it is not enough in order to obtain the finite on-shell value of the 
gravitational action. For this one would need the full solution, 
not just its asymptotic expansion near infinity. 

In this letter we obtain a three dimensional Einstein metric 
of constant negative curvature given any boundary metric
and a $(d{+}1)$-dimensional ($d>2$) conformally flat Einstein 
metric with conformal structure at infinity represented by an arbitrary
conformally flat metric. 
In the latter case, had we picked the flat metric as a representative 
of the boundary conformal structure, our solution 
would reduce to the standard anti-de Sitter metric 
in Poincar\'{e} coordinates, so
it is locally isometric to adS spacetime.
The interest in considering the solution 
with arbitrary conformally flat metric in the boundary
stems from the fact that in the adS/CFT correspondence 
the boundary metric becomes a source for the boundary 
energy momentum tensor, and in quantum field theory the sources
are arbitrary.  To obtain the on-shell value of the gravitational
action one still has to know the global structure of the 
solutions. We shall not delve into this issue here.
Nevertheless, we shall propose a general formula for 
the effective action, (\ref{EH2}), and show that it
has the correct transformation properties 
under conformal transformations. 

Let $X_{d+1}$ be a $(d{+}1)$-dimensional manifold with a boundary 
at infinity $M_d$. Given a metric $g_{(0)}$ at $M_d$ 
we seek for a solution of Einstein's equations,
\be \label{feq1}
\hat{R}_{\m \n} - \half \hat{R} \hat{G}_{\m \n} = \L \hat{G}_{\m \n},
\ee
such that $g_{(0)}$ is a representative of the conformal 
structure at infinity. Here 
$\L=-{d(d-1) \over 2 l^2}$ is the cosmological constant,
and throughout this article hatted quantities refer to $d{+}1$
dimensional object. Greek indices denote $(d{+}1)$-dimensional indices
and latin indices $d$-dimensional ones. The radial coordinate
will be denoted by $\r$.
We will work in the coordinate system
introduced by Fefferman and Graham \cite{FG} where 
the bulk metric takes the form
\begin{equation}
ds^2={l^2d\rho^2\over 4\rho^2}
+{1\over \rho}g_{ij}(x,\rho)dx^idx^j~~,
\label{11}
\end{equation}
where
\be \label{expans}
g(x, \r) = g_{(0)}  + \rho g_{(2)} + \rho^2 g_{(4)} + \ldots ~~.
\ee

Equation (\ref{feq1}) can be integrated exactly when the 
Weyl tensor vanishes. This is always true in three
dimensions, so it imposes no further conditions. In higher 
dimensions, however, it implies that the bulk metric is conformally flat. 
The Weyl tensor in $(d{+}1)$ dimensions is equal to
($\m=\{\r, i\}, i=0,...,d-1$)  
\be \label{weyl}
\hat{W}_{\k \l \m \n}=
\hat{R}_{\k \l \m \n} + 
(\hat{P}^{(d+1)}_{\l \m} \hat{G}_{\k \n} 
+ \hat{P}^{(d+1)}_{\k \n} \hat{G}_{\l \m} 
-\hat{P}^{(d+1)}_{\k \m} \hat{G}_{\l \n} 
- \hat{P}^{(d+1)}_{\l \n} \hat{G}_{\k \m}),
\ee
where 
\be
\hat{P}^{(d+1)}_{\m \n}(\hat{G}) = {1 \over d-1}\left(
\hat{R}_{\m \n} - 
{\hat{R} \over 2 d} \hat{G}_{\m \n}\right).
\ee
When the Weyl tensor vanishes the field equations (\ref{feq1}) 
become,
\be \label{feq2}
\hat{R}_{\k \l \m \n}={1 \over l^2} 
(\hat{G}_{\k \m} \hat{G}_{\l \n} -\hat{G}_{\l \m} \hat{G}_{\k \n}).
\ee
This implies that the space is locally isometric to anti-de Sitter 
space. In the coordinate system (\ref{11}), equation (\ref{feq2})
becomes\footnote{
Our conventions are as follows
$R_{ijk}{}^l=\pa_i \G_{jk}{}^l + \G_{ip}{}^l \G_{jk}{}^p - i
\leftrightarrow j$ and $R_{ij}=R_{ikj}{}^k$. With these
conventions the curvature of anti-de Sitter space comes out positive,
but we will still use the term ``metric of constant negative
curvature'' for anti-de Sitter metrics.} 
\bea
&&l^2 R_{ijkl}(g)=g_{ik} g_{jl}' + g_{jl} g_{ik}'
-g_{il} g_{jk}'- g_{jk} g_{il}'
-\r (g_{ik}' g_{jl}' - g_{il}' g_{jk}'), \label{1st} \\
&&\nabla_i g_{jk}'-\nabla_j g_{ik}' = 0, \label{2nd} \\
&&g''-\half g' g^{-1} g' =0. \label{3rd}
\eea

One can show that if (\ref{3rd}) holds to all orders and 
(\ref{1st}) and (\ref{2nd}) to lowest order, then 
(\ref{1st})-(\ref{2nd}) hold to all orders.
Differentiating (\ref{3rd}) 
one obtains,
\be
g'''=0.
\ee
This means that the expansion in (\ref{expans}) 
terminates at $\r^2$. One can integrate (\ref{3rd}),
\be \label{metr}
g=\left(1+{\r \over 2} g_{(2)} g_{(0)}^{-1}\right) g_{(0)} 
\left(1+{\r \over 2} g_{(0)}^{-1} g_{(2)}\right).
\ee
Therefore, to fully determine the metric we only need to know
$g_{(2)}$. There are two cases to consider: $d=2$ and $d>2$.
We start from the latter. In this case one easily obtains from 
(\ref{1st}),
\be \label{d>2}
g_{(2) ij}=l^2 P^{(d)}_{ij}(g_{(0)})= l^2 {1 \over d-2} 
\left(R^{(0)}_{ij} - {R_{(0)} \over 2 (d-1)} g_{(0) ij}\right),
\ee
where $R^{(0)}_{ij}$ is the Ricci tensor of the metric $g_{(0)}$ etc. From 
(\ref{1st}) and (\ref{weyl}) we find that the Weyl tensor
of $g_{(0)}$ vanishes. In $d>3$ this implies that $g_{(0)}$ is 
conformally flat. In this case Bianchi identities imply that 
(\ref{2nd}) is satisfied as well. In $d=3$ the Weyl tensor vanishes
identically. However, in this case the derivation of (\ref{2nd}) through 
Bianchi's does not go through. 
The tensor in the left hand side
of (\ref{2nd}) at $\r=0$ is the Weyl-Schouten tensor
(sometimes also called the Cotton tensor). The vanishing 
of the latter is a neccesary and sufficient condition for
conformal flatness in three dimensions (see, for instance, 
\cite{FG}). Thus, in $d=3$,
$g_{(0)}$ is required to be conformally flat as well.
To summarize, we have shown that in $d>2$ the metric 
(\ref{11}) with $g$ given by (\ref{metr})-(\ref{d>2})
is a conformally flat solution of Einstein's
equation, provided $g_{(0)}$ is also conformally flat.

When $d=2$ equation (\ref{1st}) only determines the trace,
\be
\Tr(g_{(0)}^{-1} g_{(2)})={l^2 \over 2} R_{(0)}.
\ee
{}From (\ref{2nd}) we get 
\be \label{d=2}
g_{(2) ij} = {l^2 \over 2}\left(R_{(0)} g_{(0) ij} + T_{ij}\right),
\ee
where $T_{ij}$ is a symmetric tensor that satisfies
\be \label{teq}
\nabla^i T_{ij}=0, \qquad \Tr(g_{(0)}^{-1} T)=-R_{(0)}.
\ee
$T_{ij}$ has three independent components and we have 
three equations. Since (\ref{teq}) involves a differential 
equation the solution will in general be non-local in 
terms of covariant expressions of $g_{(0)}$. 
One can restore covariance and locality by introducing 
an auxiliary scalar $\phi$, and integrate over it. 
$T_{ij}$ then is the stress energy tensor
of that scalar. From the trace condition we see that $\f$
is the Liouville field. The action for the Liouville theory 
(with no potential) is
\be \label{Liouv}
W_L= {1 \over 48 \p}
\int_{X_2} d^2x \sqrt{g_{(0)}} 
\left( \half g_{(0)}^{ij} \nabla_i \f \nabla_j \f + \f R_{(0)} \right). 
\ee
The occurrence of the energy-momentum tensor of the Liouville 
field in $g_{(2)}$ has been recently reported in \cite{BEHS}. 
The field equation for $\f$ is 
\be \label{Lieq}
\Box \f 
= R_{(0)}.
\ee
Integrating out $\f$ the action (\ref{Liouv}) 
reduces to the Polyakov non-local effective action.
The stress energy tensor is given by
\be
T_{ij} = \half \nabla_i \f \nabla_j \f + \nabla_i \nabla_j \f 
-\half g_{(0) ij} \left(\half (\nabla \f)^2 
+2 \Box \f\right).
\ee
One may check that the stress energy tensor satisfies (\ref{teq}),
provided (\ref{Lieq}) is satisfied.
One could, more generally, consider the Liouville field with a
potential $V(\phi)=\m e^{-\f}$. In this case, however, 
$\f$ becomes interacting and it is no longer auxiliary.
Furthermore, one can add to $T_{ij}$ the energy-momentum 
tensor of arbitrary conformal matter (i.e. matter with traceless,
covariantly conserved energy momentum tensor). 

According to the prescription in \cite{Gubs,Wit} the effective
action (or more properly, the generating functional of connected graphs)
of the boundary theory is given by the on-shell value 
of the action functional
\be \label{actfun}
S[\hat{G}]={1 \over 16 \p G}
\left[\int_{X_{d+1}} d^{d+1}x \sqrt{\det \hat{G}} \left(\hat{R} + 2 \L \right)
+\int_{M_d} d^d x \sqrt{\det \tilde{g}} 2 K \right],
\ee
where $K$ is the trace of the second fundamental form and $\tilde{g}$
is the induced metric at the boundary. The expression in 
(\ref{actfun}), however,
is infrared divergent. To regulate the theory we cut-off the bulk integral 
to $\r > \e$ for some cut-off $\e>0$ and evaluate the boundary 
integral at $\r=\e$. In \cite{HS} the divergences were calculated
and shown that can be removed by adding covariant counterterms.   
The focus here will be on the finite part. An important issue 
is what is the range of the $\r$ coordinate. Intuition from 
spaces that are topologically a ball suggest 
to integrate from the boundary till the center of the 
ball. In the center of the bulk, the bulk volume element should vanish.
We therefore propose to integrate $\r$ from $\e$ till 
$\r_{cr}(x)$, where $\r_{cr}(x)$ satisfies 
\be \label{eqr}
\det A(x,\rho_{cr}(x) )=0, \qquad A^2(x,\r)=g^{-1}_{(0)}g.
\ee
In the solutions we are 
considering here, $A(x,\r)$ is linear in $\r$, so $\det A$
is a polynomial of degree $d$, and (\ref{eqr}) has $d$ roots.
It is an important question which root one should pick.
Intuitively, one might expect that one should pick the 
smallest positive $\r_{cr}(x)$ (recall that the boundary is at $\r=0$). 
Furthermore, one could also demand that the area of the 
surface $\r=\r_{cr}(x)$ vanishes. This imposes conditions
stronger than (\ref{eqr}). In the case the boundary metric
is stationary, the latter condition is satisfied if the bulk lapse and 
shift functions vanish at $\r=\r_{cr}(x)$.   
In general, we expect the choice of the root to also 
depend on global issues. Leaving a detailed analysis
for future study, we now show that for any choice 
of root, the so-defined effective action transforms
correctly under conformal transformation, i.e.
it is conformally invariant when $d$ is odd,
and its conformal transformation yields
the Weyl anomaly of the boundary theory when $d$ is even. 

The renormalized effective action, i.e. the on-shell
value of (\ref{actfun}) with the infinities cancelled
by counterterms, is given by 
\begin{equation}
W_{fin}={1\over 16\pi G}{d\over l}\int d^dx \sqrt{g_{(0)}}
\int^{\rho_{cr}(x)}
{d\rho\over \rho^{d/2+1}}\det A(x,\rho )~~.
\label{EH1}
\end{equation}
We now show that it  has the correct transformation 
properties under conformal transformations provided $g_{ij}(x,\r)$
is a power series in $\r$. 
The determinant of $A(x,\rho )$ can be presented as power series
\begin{equation}
\det A(x,\rho )=\sum_k a_k(x)l^{2k}\rho^k~~, \qquad a_0=1~~,
\label{series}
\end{equation}
containing possibly infinite number of terms. For the solution 
in (\ref{metr}), where the expansion in $\r$ terminates 
at $\r^2$, $A(x, \r)$ is linear in $\r$, 
the sum in (\ref{series}) is finite and in sufficiently 
low dimensions one can explicitly obtain the roots of the 
equation $\det A(x,\rho_{cr}(x) )=0$. Feeding back in (\ref{EH1})
one then obtains the effective action,
\begin{equation}
W_{fin}={1\over 16\pi G}{d\over l}\int d^dx \sqrt{g_{(0)}}
\left( \sum_{k\neq {d\over 2}}{1\over k-{d\over 2}}\rho_{cr}^{k-{d\over 2}}(x)l^{2k}
a_k(x)+a_{d\over 2}(x)\ln\rho_{cr}(x)\right) 
~~,
\label{EH2}
\end{equation}
where the last term appears only when $d$ is even.
This expression is in general 
non-analytic in the curvatures. The starting point in the 
derivation of (\ref{EH2}) was (\ref{actfun}) which is 
the string theory effective action 
to lowest order in derivatives. Therefore, only  
the leading part in derivatives of (\ref{EH2}) 
can be reliably considered as the effective action 
of the dual conformal field theory. 

{}From the form of the field equation in (\ref{1st})-(\ref{3rd}) 
follows that in the expansion of the metric, $\r$ is always accompanied by 
a factor of $l^2$. Then, by dimensional analysis
the coefficients $a_k(x)$ in (\ref{series})
contain $2k$ derivatives.
It follows that under the infinitesimal conformal transformation
$\delta_\sigma g_{(0)}=2\sigma (x) g_{(0)}$ 
they transform as
\be \label{atransf}
\delta_\sigma a_k(x)=-2k\sigma (x) a_k(x)+\co(\nabla \sigma (x))~~.
\ee
Using this transformation property we now compute the variation
of the functional (\ref{EH2})
under a conformal variation of the boundary 
metric $g_{(0)}$. The variation of $\rho_{cr}(x)$ in $W_{fin}$
does not contribute as one sees from (\ref{EH1}): 
the corresponding contribution
is proportional to $\det A(x,\rho_{cr}(x))$  which vanishes due to 
the definition of $\rho_{cr}$. 
Performing the remaining variations we get for any $d$
\begin{equation}
\delta_\sigma W_{fin}=-{1\over 16\pi G} {2d\over l} 
\int d^dx \sqrt{g_{(0)}} \sigma(x) \left[ \rho^{-d/2}_{cr}(x)
\left(\sum_{k\neq {d\over 2}} a_k(x) l^{2k} \rho^k_{cr}(x) \right)~~ +
\nabla_i J^i\right],
\label{conf2}
\end{equation}
where the term $\nabla_i J^i$ comes from 
the terms proportional to a derivative of the $\s$ upon partial 
integration. Such contribution to the effective action can come from 
covariant counterterms \cite{HS}. The opposite
is also expected to be true, i.e. that all contributions of the 
form $\nabla_i J^i$ can be cancelled by covariant counterterms,
although, to our knowledge, this has not been shown in general.
Furthermore, these counterterms would be higher derivative terms,
and as we argued above our considerations are to leading order 
in derivatives. 
With these caveats, we proceed by omitting these terms.
 
When $d$ is odd, the terms inside the parenthesis in (\ref{conf2}) 
are just equal to $\det A(x, \r)$ evaluate at 
$\rho=\rho_{cr}(x)$. Hence,
the variation (\ref{conf2}) vanishes and the effective action is 
conformally invariant. When $d$ is even 
the infinite sum in (\ref{conf2}) reduces to single term 
$-a_{{d\over 2}}(x) l^d\rho^{d\over 2}_{cr}(x)$
and (\ref{conf2}) reads
$$
\delta_{\sigma} W_{fin}={1\over 16\pi G}dl^{d-1} \int d^dx \left( 2\sigma(x)  
a_{d\over 2}(x)\right)~~.
$$
This results in the conformal anomaly
\begin{equation}
\ca={1\over 16\pi G} 
dl^{d-1}  a_{d\over 2}(x)~~.
\label{conf3}
\end{equation}
which coincides with the result in \cite{HS}.

The effective action we have computed is, by the adS/CFT correspondence,
the effective action of some conformal field theory at strong coupling. 
Unless there is a non-renormalization theorem, one does not expect
this strong coupling result to agree with a weak coupling 
computation (using, for instance, heat kernel methods)
on the conformal field theory side. However, when $d=2$
the effective action is determined, up to a constant
which is the central charge, from the conformal anomaly.
Since there is only one possible conformal anomaly in $d=2$ 
the effective action in weak and strong coupling 
can at most differ by the value of the central charge
at weak and strong coupling. Therefore, for $d=2$ 
the effective action in (\ref{EH2}) should reduce
to the Polyakov action (or, at least, to an action which is 
in the same universality class with the Polyakov action). 
In higher even boundary dimensions, 
one can similarly obtain the effective action 
up to conformally invariant terms by integrating the 
conformal anomaly. 
Since our boundary metric is conformally 
flat, the result is expected to be, up to higher
derivative terms, equal to the effective action coming from 
integrating the Euler part of the conformal anomaly.
When the boundary dimension is odd, and since our 
boundary metric is conformally flat, the 
(conformally invariant) effective action 
reduces to a constant (which can be obtained by 
evaluating the on-shell gravitational action for 
a flat boundary metric).

In the remainder we restrict ourselves to $d=2$ and consider
the case the boundary metric $g_{(0)}$ is a general static metric,
\begin{equation}
ds^2=h(x)d\tau^2+{dx^2\over h(x)}~~,
\label{3.2}
\end{equation}
for arbitrary $h(x)$. This, for suitable $h(x)$, can be the metric 
of any two dimensional compact Riemann surface 
since the latter can always be equipped with a metric
of constant scalar curvature. Specifically, for a sphere of radius $a$, 
$h(x)=1-x^2/a^2$, for a torus, $h(x)=1$, and for a hyperbolic 
plane of radius $a$, $h(x)=x^2/a^2 -1$. One can also consider
metrics describing non-compact spaces, such as 
the metric of a 2d Schwarzschild black hole, $h(x)=1-a/x$.
The bulk metric is still a solution of Einstein's 
equation, but one has to be careful with what the 
boundary of spacetime is. Since the two-dimensional 
slices are non-compact the true boundary generically has components 
in all slices. 

Given the boundary metric in (\ref{3.2}) one can explicitly 
integrate (\ref{Lieq}) once,
\be \label{fpr}
\f=-{4J\over l^2 B} \t + \f_0(x)~~, \qquad \f_0'={h'-B \over h}~~,
\ee
where the prime indicates differentiation with respect to $x$ 
and $J$ and $B$ are integration constants. 
$g_{(2)}$ is then given by a local expression in terms of $h$
and its derivatives. Inserting in (\ref{metr}) we obtain  
\bea
ds^2 &=&
{l^2d\rho^2\over 4\rho^2} 
+{1 \over \r}
\left[h\left( (1 + {l^2 \over 16}{h'{}^2- b^2 \over h}\r)^2 + 
\r^2 ({J \over 2 h})^2\right) d\tau^2 \right. \label{metric} \\
&+&\left.
2 \r {J \over h}\left(1 + {l^2 \over 8} h'' \r\right) dx d\t 
+{1 \over h}\left( (1 +{l^2 \over 4} h'' \r
- {l^2 \over 16}{h^{' 2}- b^2 \over h}\r)^2 + \r^2 ({J \over 2 h})^2\right)
dx^2 \right], \nonumber
\eea
where $b^2=B^2 -(4J/B l^2)^2$. This metric is the most general metric
of constant negative curvature with a static metric on the boundary. 
It depends on two arbitrary constants, $b$ and $J$.

Consider the case $h(x)$ has a simple zero at some point $x=x_+$. 
This can happen 
for instance if the metric (\ref{3.2}) describes a compact closed
manifold with a topology of a sphere (in which case it vanishes
at two points), or if the metric describes a two dimensional black 
hole. Regularity of the Euclidean two-dimensional metric requires
that $\t$ is periodic with period equal to the inverse Hawking 
temperature, $T_H^{-1}=4 \p/h'(x_+)$. In this case, we find that 
in order for the bulk metric to be smooth, $J$ should be equal to 
zero, and $b=h'(x_+)$, i.e. the integration
constant $b$ is related to the Hawking temperature of the 
boundary geometry. One further finds that the latter 
condition ensures that $\f$ is regular at $x=x_+$. 
If $h(x)$ does not have a zero 
then $J$ may be non-zero. In particular, if the 
boundary geometry is a torus, so $h(x)=1$, the solution 
(\ref{metric}) is the rotating BTZ black hole solution
with mass and angular momentum given by 
$M_{BTZ}=b/2$ and $j_{BTZ}=J l$.

Let us calculate the effective action in $d=2$. 
From (\ref{EH2}) we get 
\be \label{finit}
W_{fin}={c\over 12\pi l^2} 
\int_{X_2}\left({l^2\over 4}R \ln \rho_{cr}(x)-{2\over \rho_{cr}(x)}-
{l^2\over 4}R \right),
\ee
where we introduced $c=3 l/2 G$, which is 
equal to the value of the central charge of the strongly 
coupled CFT \cite{BH,HS}.
The equation for the critical value of $\r$, $\det A(x, \r_{cr}(x))=0$,
gives a quadratic equation. The discriminant is 
equal to 
$\D ={1 \over 8} \left(\Tr T^2 - \half  
(\Tr T)^2 \right)$,
where $T$ is the Liouville energy momentum tensor.
One may check that $\D$ is always positive. 
The two roots are equal to $\r_{cr}^{\pm}=-8/[l^2(R \pm 4 \sqrt{\D})]$.
When the boundary metric is static and $J=0$ the results 
simplify because $\D$ becomes a total square.
We restrict our attention to this case.
We further choose the root that yields zero area for the
surface $\r=\r_{cr}(x)$. 
This root can be determined by demanding that $g_{\t\t}$ vanishes
at  $\r=\r_{cr}(x)$.
The result is
$\rho_{cr}(x)={16\over l^2}{h(x)\over h'^2(x_+)-h'^2(x)}$.
The critical value $\r_{cr}(x)$ is positive provided
$h''(x)<0$ and smaller than the second root if $h'''(x)>0$.

We have argued above that (\ref{finit}) should be a multiple 
of the Polyakov action (\ref{Liouv}), with the multiplicity factor being 
the central charge of the strongly coupled CFT.  
In general, one may need field redefinitions in order to 
demonstrate this fact. In the case of static metric with $J=0$ 
such complications do not arise. 
To obtain the Polyakov action we need 
to integrate (\ref{fpr}) (with $J=0$) once more, 
since it is $\phi$ that appears in (\ref{Liouv}). 
This involves a new integration constant. 
The latter can be fixed by choosing a reference manifold $D_2$
on which the quantum state is the vacuum state and
demanding that $\f_{D_2}=0$ \cite{IFS}.
Let us first consider the case the $2d$ manifold is non-compact,
and put it on the box of size $L$. With
the choice of reference manifold in \cite{IFS}, we get  
\begin{equation}
\phi(x)=\int_x^Ldx {b - h'(x) \over h(x)} +
2\ln (2\pi T^{-1}_H h^{1/2}(L)\mu )~~,
\label{2.6}
\end{equation}
where $\m$ is related to the size of the reference manifold. 
It is straightforward to evaluate (\ref{Liouv}). The result is
\begin{equation}
W_L={1\over 12 b}\int_{x_+}^Ldx \left( {b^2-h'^2\over 2h} \right) 
-{1\over 12}\phi (x_+)~~,
\label{3.18}
\end{equation}
where $x_+$ is the zero of $h(x)$ and $\phi (x)$ satisfies (\ref{2.6}).  
One may verify that $W_{fin}$ and $c W_L$ differ only by boundary terms,
and that the latter vanish 
for asymptotically flat boundary spacetimes in the large $L$ limit
(modulo terms independent of the boundary geometry). The same 
calculation, yielding again agreement, 
can be done for the case of spherical geometry
by considering two coordinate patches. More generally, if
$W_{fin}$ and $c W_L$ only differ by boundary terms,
one can achieve exact agreement by suitable choice of the
reference manifold. Let us finally note that our results are 
in agreement with the results in \cite{marc}
where it was shown that the asymptotic dynamics
of three-dimensional Einstein gravity with constant negative curvature
is described by Liouville theory.

{\bf Acknowledgments} \\
We would like to thank J. de Boer, R. Dijkgraaf, G. Gibbons and 
R. Myers for useful discussions. KS is supported by the Netherlands
Organization for Scientific Research (NWO).

\end{document}